# Influence of homo-buffer layer on stress control of sputtered (Ba$_{0.45}$,Sr$_{0.55}$)TiO$_3$ thin films on Pt-Si

Y.K. Vayunandana Reddy[1], Guillaume Guegan[2], Mohamed Lamhamdi[3], Virginie Grimal[1] and Patrick Simon[4]

[1]LEMA, UMR 6157 CNRS-CEA, Université François Rabelais, Parc de Grandmont, 37200 Tours, France.
[2]R&D, STMicroelectronics, 16 rue Pierrie et Marie Curie, 37071 Tours, France
[3]LMP, STMicroelectronics, 16 rue Pierrie et Marie Curie, 37071 Tours, France
CEMHTI-UPR CNRS 3079, Universite d'Orleans, BP6749, 45067 Orleans Cedex, France

To engineer strain relaxation of sputtered BST thin films on Pt-Si wafers, homo-buffer layer method was applied to eliminate Pt hillock formation. Thin BST homo-buffer layers were deposited at room temperature and subsequently the main BST layer was deposited at 650°C, Pt hillock free BST films were obtained with homo-buffer thickness above 5 nm. Relatively good electrical properties were obtained for BST thin films with 15 and 25 nm homo-buffer layer (T= 30 % at 5V and tan $\delta$= 0.018).

For past few years, the fundamental properties of ferroelectric (Ba$_x$,Sr$_{1-x}$)TiO$_3$[1] thin films has extensively studied for their applications in dynamic and non-volatile random access memory devices (DRAM and NV-RAM) [2], and microwave device applications such as, phase shifters, tunable oscillators, and delay lines[3]. (Ba$_x$,Sr$_{1-x}$)TiO$_3$ can display paraelectric phase by controlling the Ba/Sr ratio with the merits of high permittivity of BaTiO$_3$ and structural stability SrTiO$_3$. (Ba$_{0.45}$Sr$_{0.55}$)TiO$_3$ (BST) is a paraelectric with Curie temperature of -50°C [4].

Platinum (Pt) has been considered as the most desirable metal electrode due to its potential properties [5]. Pt is chemically stable in an oxidizing environment and maintains its conducting properties under thermal processing. However, due to the high processing temperature (> 500°C) synthesis of ferroelectric oxide films, difference in lattice mismatch and thermal expansion coefficient (TEC) of the film-substrate leads to lattice mismatch and thermal strain (stress) paving the way for the formation of Pt hillocks are serious concern for capacitor structure. TEC of Si (2.6x10$^{-6}$/°C) is much smaller than Pt (9x10$^{-6}$/°C) and BST (10.5x10$^{-6}$/°C) [6]. Pt hillocks are formed during thermal cycling of thin films and are related to strain/stress relief process in the film [7].

In the Pt/TiO$_x$ electrode stack, the Pt hillock formation is due to the relaxation process of the compressive stress in the film; the Pt atoms move to the film surface from the interior along the grain boundary. Therefore, at the initial state of the Pt hillock, the mass for its growth is supplied by grain boundary diffusion. The diffusion of Pt atoms along the grain boundary occurred actively toward the film surface to relax the strong compressive stress. The Pt atoms on the film surface, which was diffused out from the interior, departed toward the hillock along the film surface and afterwards accumulated just around the hillock. To control the strain/stress of the BST films, it's necessary to make strain/stress relaxation process either at very beginning of the film growth or move to the larger thicknesses [8]. Mechanical misfit strain in films is driven by the competition between strain buildup due to lattice and thermal mismatches and strain relaxation, to control of the mechanical misfit strain in ferroelectric films attempts were made in the past by changing the film thickness [9].

In this present study, very thin homo-buffer layer (BST) was used to engineer the strain relaxation process to control the stress build up from substrate stack and BST thin film and eventually to eliminate the Pt hillock formation. This method allows us to change only the homo-buffer layer thickness and keep same the main BST film deposition conditions and substrate. In this way, first very thin BST homo-buffer layers of 5, 10, 15 and 25 nm were grown on the substrate at room temperature (RT). At this stage the buffer layer is nearly amorphous, introduces a super-saturation of point defects, which increases the number of nucleation sites on the surface, which has an impact on strain relaxation of the BST film being subject to the substrate. In the second step, substrate is heated up to the growth temperature of conventional BST deposition temperature (650°C) and keep for some time (annealing), this annealing at high temperature will lead to an increase of the crystalinity of the first layer. In addition, existing misfit dislocations move to the nearby film/substrate interface, which leads to high degree of relaxation from the lattice mismatch between film and substrate with smooth surfaces. Conventional BST layer will deposit on the relaxed BST homo-buffer layer. In this letter, we report the effect of the homo-buffer layer on strain relaxation and permittivity of the BST films.

Thin films of BST thin films were deposited on commercially available 6 inch Pt coated Si wafers (Pt/TiO$_2$/SiO$_2$/Si, here on Pt-Si) by rf magnetron sputtering with a stoichiometric BST ceramic target (Dia. 150 mm). All BST thin films were deposited with sputter power of 250 W. Thin homo-BST buffer layers (5 to 25 nm) were deposited under pure Ar=100 sccm (13.5 mTorr) at room temperature (RT) and main BST thin films were deposited under 14.2 mTorr (Ar/O$_2$=100/10 sccm) pressure at 650°C on top of the buffer layer. Once the thin film deposition is over, wafers were cooled down to RT under 3.6 mTorr O$_2$ pressure. For electrical measurements 500 nm thick Pt top electrodes (A=0.3 mm²) were deposited at room temperature by dc magnetron sputtering.

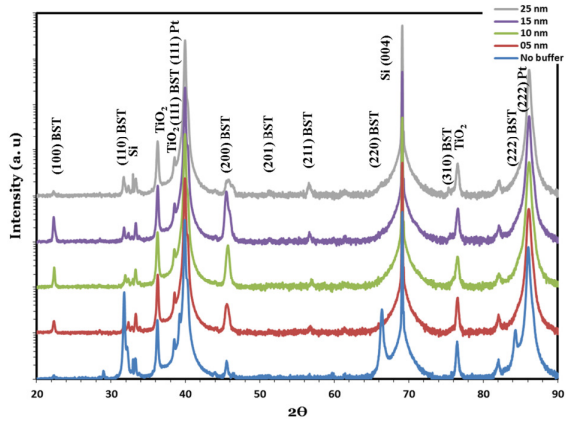

**Fig. 1 X-ray diffractograms of BST thin films**

The structural properties were characterized by Bruker D8 Discover high resolution x-ray diffractometer (HR-XRD) using Cu Kα radiation (1.5406 Å) operating at 40 kV and 40 mA. To obtain surface roughness and grain size Veeco Instruments Dimension 3100 SPM atomic force microscope (AFM) was used in tapping mode. Hitachi scanning electron microscopy (SEM) was used to study the film surface in large area. Raman spectroscopy characterization was performed by Renishaw Invia Reflex device with $Ar^+$ laser excitation radiation 514.5 nm with the laser power of few mW. Capacitance-voltage (C-V) measurements were done with sweeping of 0 to 5V in a metal-insulator-metal (MIM) configuration by HP 4294A impedance analyzer at 100 kHz. SIMS depth profiles were carried out with a CAMECA IMS 4F6 instrument using a Cs+ primary beam.

X-ray diffraction patterns of the BST thin films were recorded and depicted in Fig. 1. BST thin films were polycrystalline and showing two types of trends: textured and randomly oriented. Films with no buffer layer were strongly textured in (*hk0*) [i.e., (110), (220), and (330)] orientations and with buffer layer they were randomly orientated, but with weakly (*h00*) textured. The observed variation in the crystallographic nature of their evolved phases is due to the difference of buried homo-buffer layer's nature. The above results imply that particularly, the thicknesses of the underlying thin BST buffer layer dictates phase transformation kinetics and, hence the nucleation and growth of the perovskite BST. It also appears that the crystallographic nature of the underlying BST homo-buffer layer have strong influence on the development of perovskite BST structure. Lattice constant of the main BST is increasing with increasing homo-buffer layer thickness. Strain relaxation was occurred at the 10 nm homo-buffer layer thickness.

Optical microscope and SEM was used to monitor the changes of the surfaces of homo-buffer layer and main BST (Fig. 2). Pt hillocks were observed on BST film with no buffer layer, by inserting a thin 5 nm homo-buffer layer hillocks density was reduced and with 10 nm homo-buffer hillocks were disappears on the main BST surface. It is clear that homo-buffer layer effectively controlled the Pt hillock formation, but cracks were appeared. The size of the Pt hillocks were much smaller (~130 nm) compared to the reported in the literature [10]. From AFM scans the grain size and average roughness of BST films were estimated for with and without homo-buffer layers. It's noticed that grain size (65 nm) and roughness (~ 3.5 nm) was larger for BST film without homo-buffer layer compared to the with 15 nm buffer layer which has 50 nm and 3.3 nm, respectively. During the annealing (heating), existing dislocations of the homo-buffer layer move to nearby film/substrate interface and new dislocations are also formed at the interface, which lead to effective strain relaxation from the lattice mismatch between the film and the substrate. Such fast relaxation at 10 nm is probably due to the high densities of dislocations and nucleation sites in the first layer which have low crystallinity (quasi-amorphous) as it was deposited at RT. Annealing of this quasi-amorphous layer leads crystallization of the perovskite phase takes place in equilibrium with misfit dislocation formation, resulting in a completely relaxed template (stress free) for the subsequent continuation of growth. The relaxation mechanism of with or without buffer layer is, therefore, fundamentally different from that of the normally grown film at high temperature.

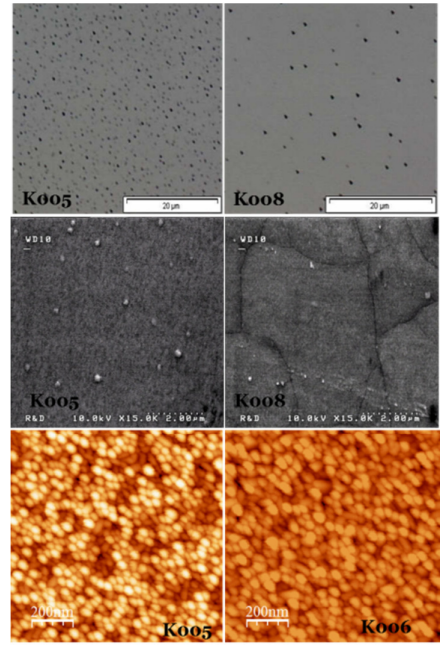

**Fig. 2 Optical, SEM and AFM images of BST**

In order to understand any inter-diffusion of layer stack, SIMS analysis was performed on films and no inter-diffusion was observed for any of the films. The well-known reported penetration of underlying Ti to the upper surface of Pt [11] is not the case in this study, as can be seen from the SIMS spectra (Fig. 3) of the BST without buffer layer, there is no diffusion of Ti in to Pt is observed, this result is well consistent with the observation by Waser et al [12].

Raman spectra (Fig. 4) of BST films showing four broad bands at 170, 295, 535 and 780 $cm^{-1}$ are attributed to second order Raman scattering and no first order Raman modes were observed its demonstrating that BST films were in cubic phase. Band at ~890 $cm^{-1}$ appears for all BST films except for 25 nm buffer layer, this mode seems originated because of the residual stress in the film [13,14]. This is clearly demonstrating that with 25 nm buffer layer, stress was completely relaxed and then stress mode disappears. This is correlating with microscopic image which doesn't show any hillocks.

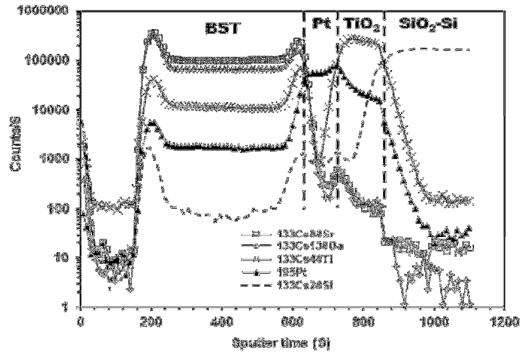

**Fig. 3 SIMS spectra of BST without buffer layer**

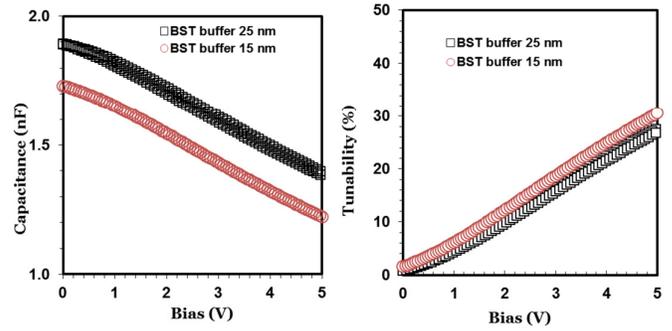

**Fig. 5 C-V and tunability of BST films**

Measurements of C-V were done at 100 kHz on all samples at room temperature and depicted in Fig.5. Capacitors without homo-buffer layer were short circuited because of the Pt hillocks. No short circuit was observed for films with homo-buffer layer thickness > 10 nm, suggesting hillock formation or stress was controlled effectively. The tunability is the variation in the capacitance at a given bias voltage (V) from its value at zero bias such that, percentage of the tunability can be written as T% = [($C_{V=0}$-$C_{V=a}$) / $C_{V=0}$*100]. BST films with 15 and 25 nm homo-buffer layers were shown good electrical properties with the tunabilities of 30% at 5V and the dissipation factor (tan δ) of 0.018 at 0V. C-V curves of BST films doesn't show any butterfly effect concluding that films are in paraelectric phase, which shows direct correlation with XRD and Raman study. Capacitance was smaller for 15 nm buffered BST film compared to the 25 nm buffered BST, it might be because of some stress is still exist in the 15 nm buffered film which is correlated with Raman mode at ~890 cm$^{-1}$, but interestingly tunability is comparatively similar for both films.

In summary, the homo-buffer layer method was applied to sputtered BST thin films deposited on Pt-Si wafers to engineer the lattice and thermal strain/stress relaxation which originate from lattice and thermal mismatch of film and substrate. By depositing a thin homo-buffer layers (> 5 nm) at RT before depositing the main BST layer at 650°C, an almost stress free BST films without any Pt hillocks were obtained. This study demonstrated that by engineering the strain relaxation process with the help of BST homo-buffer layer (>5 nm) Pt hillocks were effectively controlled. BST thin films with 15 and 25 nm homo-buffer layers were shown good electrical properties (T= 30% and tan δ= 0.018). Studies are under way to implement this homo-buffer layer method for high BST deposition temperatures (>650°C).

### References

[1] T. Horikawa, N. Mikami, T. Takita, J. Tanimura, M. Kataoka, K. Sato, and M. Nunoshita, *Jpn. J. Appl. Phys., Part 1,* **32** 4126 (1993).
[2] J. S. Horwitz; W. Chang; A. C. Carter; J. M. Pond; S. W. Kirchoefer; D. B. Chrisey; J. Levy; C. Hubert, Int. Ferr. **22** 279 (1998).
[3] Ferroelectric film devices, Handbook of Thin film devices Vo. 5 (Acad. Press, Sand Diego, 2000).
[4] A.K. Tagantsev, V.O. Sherman, K.F. Astafiev, J. Venkatesh, and N. Setter, J. Electroceram., **11** 5 (2003).
[5] G.T. Stauf, C. Ragaglia, J.F. Roeder, D. Vestyck, J.P. Maria, F.T. Ayguavives, A.I. Kingon, A. Mortazawi, A. Tombak, Integr. Ferroelect. **39** 1271 (2001).
[6] Z. –G. Ban and S. P. Alpay, J. Appl. Phys. **91** 9288 (2007).
[7] P.D. Hren, H. Al-Sghareef, S.H. Rou, A.I. Kingon, P. Buad, E.A.Irene, Mater. Res. Soc. Symp. Proc. **260** 575 (1992).
[8] L. S. –J. Peng, X. X. Xi, B. H. Moeckly, and S. P. Alpay, Appl. Phys. Lett. **83** 4592 (2003).
[9] H. Li, A. L. Roytburd, S. P. Alpay, T. D. Tran, L. Salamanca-Riba, and R. Ramesh, Appl. Phys. Lett. **78** 2354 (2001).
[10] W. W. Jung, S. K. Choi, S. Y. Kweon, and S. J. Yeom, Appl. Phys. Lett., **83** 2160 (2003).
[11] J. L. Cao, A. Solbach, U. Klemradt, T. Weirich, J. Mayer, U. Böttger, U. Ellerkmann, P. J. Schorn, P. Gerber and R. Waser, J. Am. Ceram. Soc. **89** 1321 (2006).
[12] J. L. Cao, A. Solbach, U. Klemradt, T. Weirich, J. Mayer, H. Horn-Solle, U. Böttger, P. J. Schorn, T. Schneller, and R. Waser, J. Appl. Phys. **99** 114107 (2006).
[13] H. Guo, L. Liu, S. Ding, H. Lu, Y. Zhou, B. Cheng, and Z. Chen, J. Appl. Phys. **96** 3404 (2004).
[14] S. Y. Wang, B. L. Cheng, Can Wang, S. Y. Dai, K. J. Jin, Y. L. Zhou, H. B. Lu, Z. H. Chen, and G. Z. Yang, J. Appl. Phys. **99** 013504 (2006).

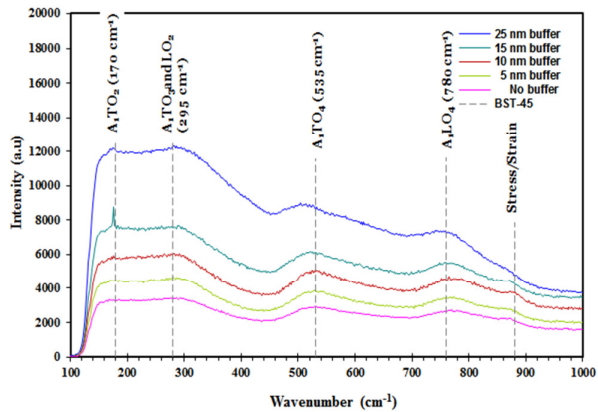

**Fig. 4 Raman spectra of BST films**